# A Neural Model of Number Comparison with Surprisingly Robust Generalization

Thomas R. Shultz
Department of Psychology and School of Computer Science, McGill University

Ardavan S. Nobandegani
Department of Psychology and Department of Electrical & Computer Engineering, McGill University

Zilong Wang
Integrated Program in Neuroscience, McGill University

## Abstract

We propose a relatively simple computational neural-network model of number comparison. Training on comparisons of the integers 1-9 enable the model to efficiently and accurately simulate a wide range of phenomena, including distance and ratio effects and robust generalization to multidigit integers, negative numbers, and decimal numbers. An accompanying logical model of number comparison provides further insights into the workings of number comparison and its relation to the Arabic number system. These models provide a rational basis for the psychology of number comparison and the ability of neural networks to efficiently learn a powerful system with robust generalization.

## Introduction

We all occasionally engage in comparing two numbers to each other to determine which is smaller or larger. Such comparisons are common, seemingly automatic, and roughly accurate. Examples include an investor estimating returns on two different investments, a shopper looking for a good deal, or a young child selecting the lesser or greater collection of similar items such as candies or toys.

Early psychological research quickly established that such number comparisons are not done by consulting a memory lookup table. That solution would be inefficient because of the number of possible numerical comparisons is potentially infinite, and the inability of table lookup to account for subtle characteristics of the numbers being compared. For example, a sample of female students at Stanford University asked to quickly and accurately select the larger of two single-digit integers in the range 1-9 took longer to decide and made slightly more errors when the difference between the two integers was small than when it was large (Moyer & Landauer, 1967). The authors noted that such phenomena also occurred when people were asked to compare physical quantities such as length, weight, or loudness, suggesting that magnitudes had to be estimated and compared before a decision could be made.

Subsequent experiments found that the ratios between the two integers created similar phenomena for both reaction time and errors (Adriano, Girelli, & Rinaldi, 2021). When the smaller/larger ratio of the two integers was large (e.g., 8/9 = .889), reaction time and errors increased as opposed to comparisons involving smaller ratios (e.g., 2/3 = .667). Summarizing these empirical findings, integer comparisons are quicker and more accurate when the relevant magnitudes are easier to distinguish from each other.

Eventually, various computational models were formulated to help explain these and other related psychological phenomena. Virtually all these models utilized artificial neural networks. There are currently at least 15 distinct associated empirical phenomena to explain, and no model that simulates and explains all of these phenomena, although one model comes close (Huber, Nuerk, Willmes, & Moeller, 2016).

For these and other reasons, it is clear that number comparison is still not fully understood. Importantly, there has been no systematic study of the early learning and development of number comparison ability and whether and how it might enable later-developing related skills involving multidigit integers, negative integers, and decimal numbers. Here, we present a novel computational model of number comparison that simulates the important difference and ratio effects in both reaction time and error for digits 1-9. The model also exhibits remarkable generalization ability to these advanced number types, despite its relative simplicity compared to other existing models.

## Method

To simulate learning how to compare numbers, we use an artificial neural network algorithm called Cascade-correlation (CC) (Fahlman & Lebiere, 1990; Shultz, Nobandegani, & Fahlman, 2022). An artificial neural network includes a set of units (representing neurons or groups of neurons) that can be variously active, where activity is an average firing rate over a particular time period, and connection weights (representing synapses), whose initial values are small and random, but which change during learning in order to reduce sum-of-squared-error in network outputs. Such networks learn to discriminate categories of the examples that they are trained on. Activations are passed forward, from input units that describe example pairs to output units describing the response to that input. Network output can be considered a prediction of what should appear at the output layer, while target output represents what does actually happen at the output layer. During learning, connection weights are adjusted so that overall network error is reduced:

$$E = \sum_o \sum_p (A_{op} - T_{op})^2 \quad \text{(Equation 1)}$$

where $E$ is sum-of-squared error, $A$ is the actual output activation for unit $o$ and pattern $p$, and $T$ is the target output activation for this unit and pattern.

CC training starts with a two-layer network (i.e., only the input and the output layer), and then recruits hidden units one at a time, if needed, to solve the problem being learned. Number comparison turns out to be sufficiently simple that there is no need to recruit any hidden units when the 1-9 integers are being compared. Our CC networks use an asymmetric sigmoid activation function:

$$y_i = \frac{1}{1+e^{-x_i}} \quad \text{(Equation 2)}$$

where $y$ is the receiving unit $i$'s output, $x_i$ is the net input to unit $i$, and $e$ is the exponential function. Thus, output unit activations range from a floor of 0 to a ceiling of 1.

A drawing of our number comparison (NC) network is presented in Figure 1. As is common in neural networks, input values are specified numerically. For example, to describe a comparison of the numbers 2 and 3, activity of the left input unit is 2 and that of the right input unit is 3. The activation signal sent to the output units is the sum of products of each sending unit activation and its connection weight, passed through the activation function specified in Equation

2. A bias unit is typical in neural networks, functioning like the intercept added into a linear equation. The bias unit has a constant input of 1 and learnable output connection weights. This allows shifting the activation function in binary comparisons to the left or right, which may be important for successful learning.

A chronic problem in neural networks is an inability to simulate variation in reaction time because activation may pass through the network in constant time in single-layer networks like ours. A common fix is to add a constraint satisfaction module to the output of the neural network that elaborates the network's final decision by cycling the network output to decision nodes in a gradual fashion, while recording the number of cycles required for a decision to satisfy a specified constraint. In our case, these two decision weights, fixed at 0.1, convey the magnitude information to corresponding decision units that are initialized to 0. When the absolute difference between the decision units reaches a value of .9, cycling stops and the number of cycles reached is reported as an index of reaction time. The more decisively different the magnitude estimates are, the less time it takes to satisfy this criterion.

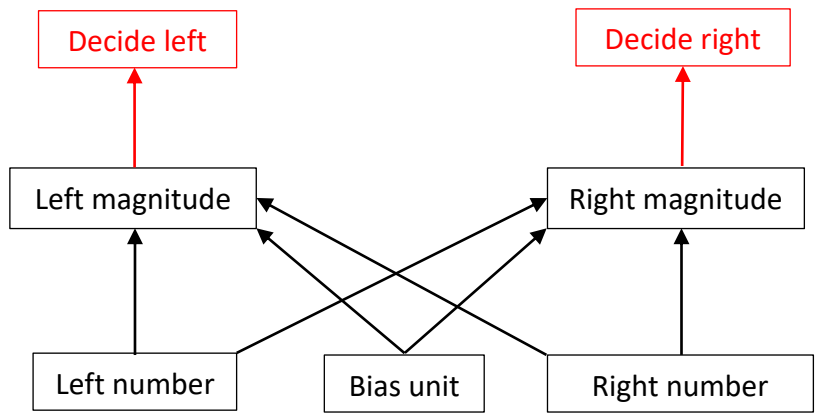

Figure 1. Diagram of our Number Comparison (NC) model. Magnitude outputs of the CC learning network (here, colored black) are fed to the decision module (colored red) which eventually generates an explicit decision and reports the number of time cycles required to reach that decision. The seven units in the CC network (represented here as rectangles) are variously active and send their activity signals over connection weights (black arrows), exciting or inhibiting activity in receiving units. The six connection weights in each CC network are learned in the service of reducing sum-of-squared-error in the magnitude estimations, while the two connection weights in the decision module (red arrows) are each fixed at 0.1 to implement a gradual buildup of decision strength.

## Results

Simulation results focus on two important empirical effects (difference and ratio) and eventual generalization to later developing skills (with multi-digit numbers, negative integers, and decimal numbers). We run 20 NC networks in each simulation, allowing statistical analysis of the results.

**Difference effect**

The difference and ratio effects are evident in measures of both error and reaction time. Figure 2 shows that error decreases as a function of the absolute difference between each pair of integers being compared, reflected as a large main effect in a repeated-measures ANOVA, $F(7, 133) = 191$, $p < .00001$, $\eta_p^2 = .91$. Tukey comparisons between adjacent means are statistically significant at differences from 1-6, but not from 6-8, where error is virtually 0. Each of four simulations involves 20 networks, each with small, randomly initialized connection weights in the learning module.

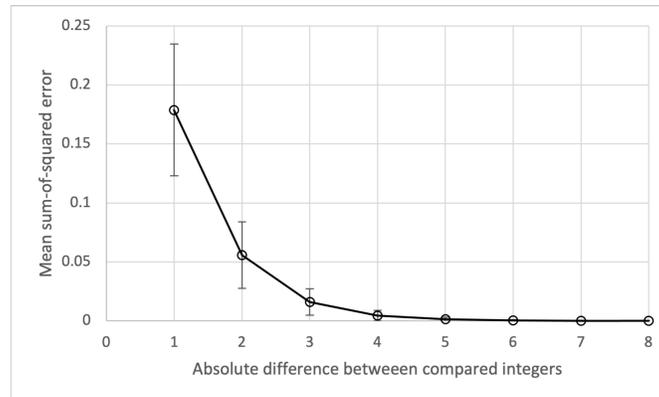

Figure 2. Mean error (with SD) as a function of difference between the two integers being compared.

In this simulation, networks learn in a mean of 9.65 epochs, with an SD of 4.20. Each epoch presents each of the 72 paired-integer comparisons once, along with the correct feedback about which is larger. At the end of each epoch, the 6 connection weights in the learning network are adjusted to lower the sum-of-squared error.

Two examples of learning progress by individual networks are shown in Figure 3. The network on the left starts with no pairs correct and reaches complete correctness in 7 epochs. The network on the right starts with half of the pairs correct (by chance) and reaches fully correct performance in 9 epochs. Such up and down progress is typical in this learning. In CC, initial connection weights are initialized to small random values.

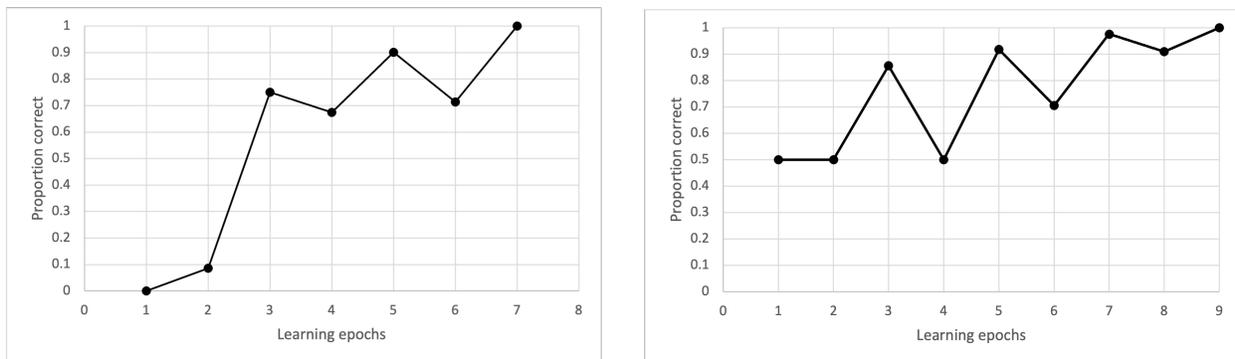

Figure 3. Learning progress in each of two representative networks across epochs.

In an analogous ANOVA of cycles to reach a decision, Figure 4 shows that mean cycles to reach a decision decrease as a function of the absolute difference between the integers being compared, with a large main effect of difference, $F(7, 133) = 158$, $p < .00001$, $\eta_p^2 = .89$. Tukey comparisons between adjacent means are statistically significant at differences from 1-7, but not from 7-8, where cycles approach a floor asymptote.

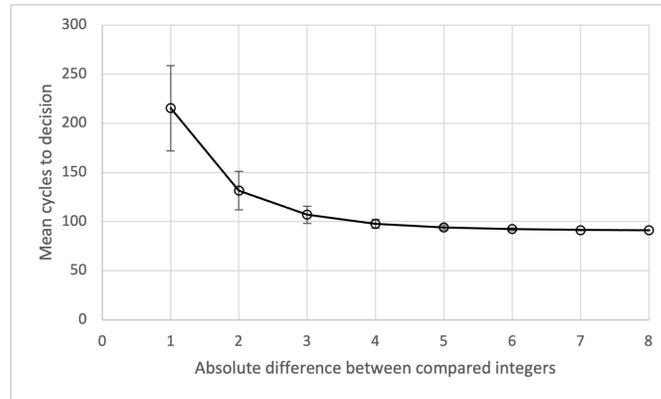

Figure 4. Mean cycles to reach a decision as a function of difference between the two integers being compared, in 20 neural networks.

In a later Results section, we develop a logical model of our NC system to gain insights into how and why this neural network accomplishes accurate number comparison. An interesting prediction of the logical model is that the positive and negative weights entering an output node are additive inverses of each other, ensuring that they sum to about 0. Other aspects of the learned connection weights vary randomly, as initial weight values are set randomly.

We test this additive inverse prediction by analyzing the sizes and signs of the learned connection weights. A typical example of this analysis is shown in Figure 5, which shows the final connection weights learned by a single representative network in the number-difference simulation. There, the weight pairs that become approximate additive inverses are:

- LN-LM (1.4) and RN-LM (-1.22). These are the two weights multiplying the left and right integers leading to the **left** magnitude output, which for this network are 1.4 and -1.22, respectively.
- RN-RM (.97) and LN-RM (-1). These are the two weights multiplying the left and right integers leading to the **right** magnitude output, which for this network are .97 and -1, respectively.

Direct weights (that favor left side or right side) become positive, while crossover weights (from one side to the other side) become negative, which helps to inhibit the opposite alternative. This connection-weight pattern is approximated by all the networks in all the present simulations, although every network solution is effectively unique because they each start with randomly different initial weight values.

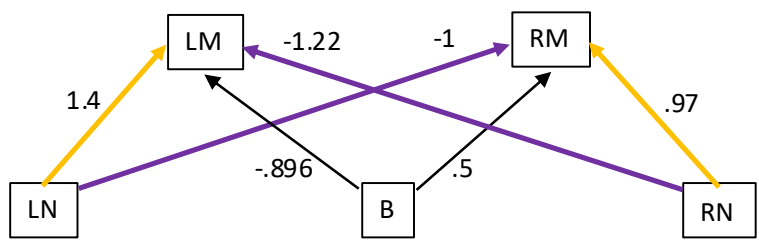

Figure 5. Connection weights in a representative single network in the difference-effect simulation. LN is left input number, B is bias unit, RN is right input number, LM is left magnitude output, and RM is right magnitude output. Positive input weights are colored gold, negative input weights are colored purple, and bias weights are colored black.

The additive-inverse prediction of the logical model is confirmed in Figure 6 which plots the value of the crossover weight as a function of the direct weight entering a magnitude output unit. Linear regressions explain nearly all the $R^2$ variance, as noted in Figure 6. Degree of spread along the $x = y$ line is governed by the weight from the bias unit, which is always on with an input of 1, regardless of the varying numerical inputs across the 72 integer pairs of digits 1-9.

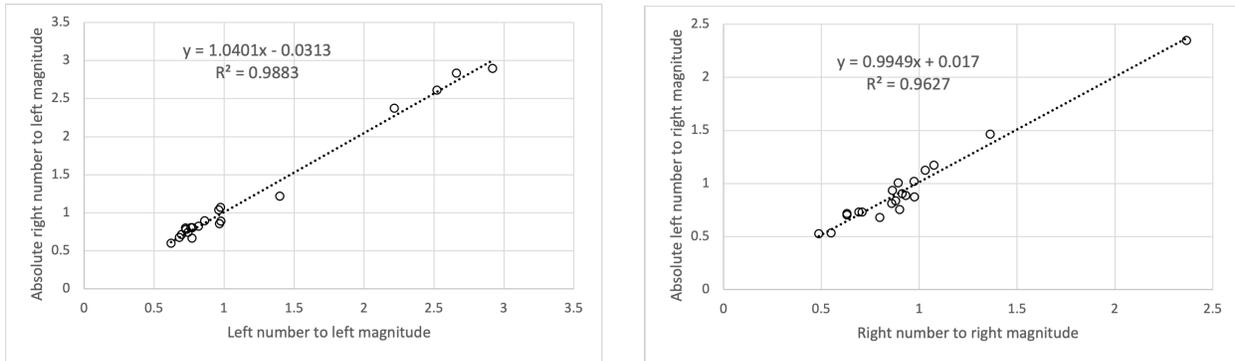

Figure 6. Crossover weights as a function of direct weights for each of 20 networks in the number-difference simulation. Each circle represents a pair of weights entering the left or right magnitude output unit. The fact that these weights cling to the $x = y$ line shows that the two weights are additively inverse, summing to approximately 0 and confirming the prediction of the logical model described in a later section.

**Ratio effect**

A second simulation addresses the ratio effect, the idea that increasing ratios of the smaller to larger integer would increase difficulty in the form of both increases in error and slower reaction time. Figure 7 presents the impact of these ratios on error. A repeated-measures ANOVA of error yields a large main effect of ratio, $F(26, 494) = 92$, $p < .00001$, $\eta_p^2 = .83$. Sum of squared error increases with the ratio of the smaller to the larger integer being compared.

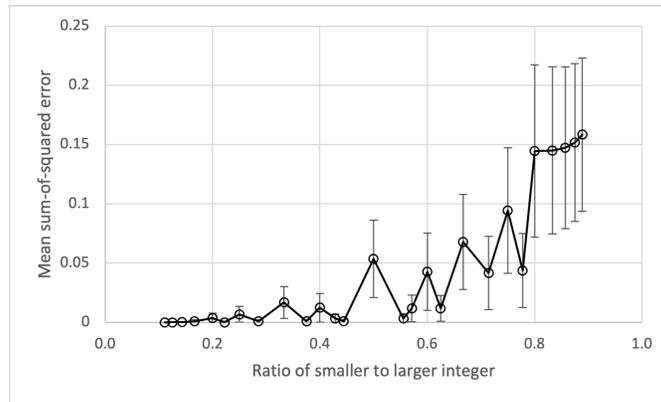

Figure 7. Mean error as a function the ratio of the smaller to larger integer being compared.

Similarly, an analogous ANOVA produces a large main effect of ratio on mean cycles to make a decision, $F(26, 494) = 83$, $p < .00001$, $\eta_p^2 = .81$. As shown in Figure 8, it takes longer to reach a decision as the ratio of smaller to larger integer increases.

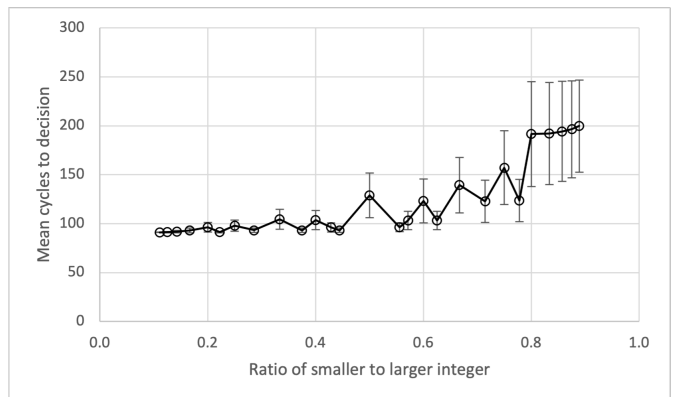

Figure 8. Mean cycles to reach a decision as a function of integer ratio.

Although it was noted at the dawn of number comparison research that distance and ratio measures could be related (Moyer & Landauer, 1967), we precisely quantify this relationship by computing a Pearson correlation between absolute differences and smaller/larger ratios for the integer pairs 1-9, $r(34) = -.843$, $p < .00001$. There are 72 of these unequal pairs, but only 36 if the direction of the differences is correctly ignored. This strong relationship implies that difference and ratio effects could be well confounded in some studies of the 1-9 integer pairs, although the distance to a perfect correlation (.157) leaves room for the two effects to be somewhat independent in this integer range.

**Generalization to advanced number types**

Three additional simulations examine whether training only on pairs of the 9 single digits would generalize to novel combinations of two, three, and four digits. The answer is an emphatic and somewhat surprising *yes*. Each of these simulations uses 20 randomly drawn pairs of 2-, 3-, or 4-digit numbers. As shown in Table 1, training on the 1-9 single digits generalizes strongly to untrained novel comparisons of 2-, 3-, and 4-digit integers.

In the last row of Table 1 are the potential numbers of digit pairs to which this level of generalization applies. Although the NC model could generalize to untrained integers of more than 4 digits, we do not have access to a computer that could generate the potential number of patterns at 5 digits. Suffice it to say that this model generalizes robustly from learning the 72 pairs of the first 9 integers.

Table 1. Mean proportion of correct generalizations out of 20 randomly selected example pairs, with SD and number of possible permutations of those digits.

|  | 2 digits | 3 digits | 4 digits |
|---|---|---|---|
| Integer range | 10-99 | 100-999 | 1000-9999 |
| Mean correct | .97 | .96 | .93 |
| SD correct | .04 | .02 | .08 |
| Potential number of pairs | 8,010 | 809,100 | 80,911,000 |

Two further simulations show similarly strong generalization from learning the 1-9 pairs to untrained negative integers and 2-digit decimal numbers (Table 2).

Table 2. Mean proportion of correct generalizations to 72 test pairs of negative numbers (-1 to -9) and 72 test pairs of decimal numbers (1.5 to 9.5), each increasing in steps of 1.

| Indicator | Negatives | Decimals |
|---|---|---|
| Mean correct | .90 | .9993 |
| SD correct | .07 | .003 |

**Logical Neural-Network Models of Number Comparison**

To provide additional insights into how and why the NC neural-network simulation model works, here we develop a logical model of number comparison. We start with a definition of the number comparison problem. Given two input numbers *num1*, *num2* decide whether *num1* > *num2*, *num1* = *num2*, or *num1* < *num2*.

The logical model rests on a fundamental property in linear algebra, illustrated in Equations 3-5, assuming that $\beta$ is a positive real-valued number:

$$num1 > num2 \equiv (num1 - num2) > 0 \equiv \beta(num1 - num2) > 0 \quad \text{(Equation 3)}$$
$$num1 = num2 \equiv (num1 - num2) = 0 \equiv \beta(num1 - num2) = 0 \quad \text{(Equation 4)}$$
$$num1 < num2 \equiv (num1 - num2) < 0 \equiv \beta(num1 - num2) < 0 \quad \text{(Equation 5)}$$

In preparation for a dual-output logical model to match our computational neural network NC model, we first develop a simpler, single-output logical model, shown in Figure 9. Here, $\sigma(\cdot)$ is a symmetric sigmodal activation function, $\sigma(x) = \frac{1-e^{\alpha x}}{1+e^{\alpha x}}$ where $\alpha$ is a positive number which controls the steepness of the sigmoidal curve. $\beta$ is a positive real-valued weight. Importantly, Figure 9 implies a family of logical models, each of which satisfies the condition that the two weights entering the output unit should sum to 0.

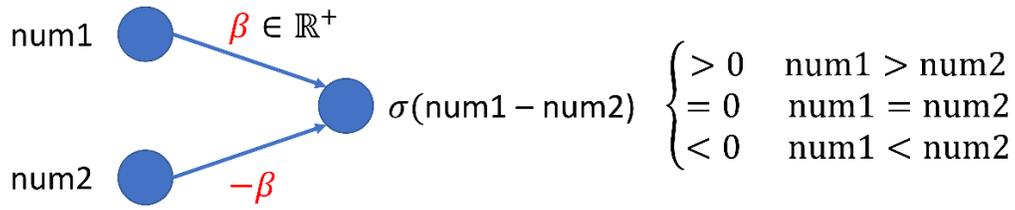

Figure 9. A logical model of number comparison with a single output unit.

We next extend this single-output logical model in Figure 9 to have two outputs, in conformity with our computational neural-network NC simulations in the previous Results section. The dual-output extension is shown in Figure 10, where $\alpha$ and $\beta$ are positive, real-valued weights. It expresses a family of dual-output networks mandating the following decisions: a) if the left output magnitude exceeds the right output magnitude, decide that the left input number is greater than the right input number; b) if the left output magnitude equals the right output magnitude, decide that the left input number equals the right input number; and c) if the left output magnitude is less than the right output magnitude, decide that the left input number is less than the right input number.

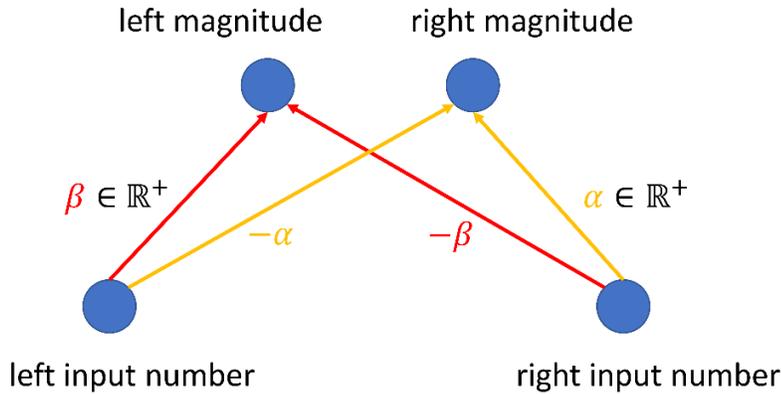

Figure 10. A logical neural-network model of number comparison with two output units.

Comparing Figures 9 and 10 reveals that the dual-output model comprises two copies of the single-output model. Specifically, the left output unit in Figure 10 receives the same weights ($\beta$ and -$\beta$) indicated in Figure 9. Similarly, the right output unit in Figure 10 receives two additive-complement weights ($\alpha$ and $-\alpha$) from the two input units. In Figure 10, the aggregate input entering the left output unit is ($\beta \times$ left input number) – ($\beta \times$ right input number), where $\beta$ is a positive, real-valued number.

More formally, Equations 6-8 govern the left output:

$$\{(\beta \times \text{left input}) - (\beta \times \text{right input})\} > 0 \equiv \text{left output} > \text{right output} \quad \text{(Equation 6)}$$
$$\{(\beta \times \text{left input}) - (\beta \times \text{right input})\} = 0 \equiv \text{left output} = \text{right output} \quad \text{(Equation 7)}$$
$$\{(\beta \times \text{left input}) - (\beta \times \text{right input})\} < 0 \equiv \text{left output} < \text{right output} \quad \text{(Equation 8)}$$

Similarly, Equations 9-11 govern the right output:

$$\{(\alpha \times left\ input) - (\alpha \times right\ input)\} > 0 \equiv left\ output > right\ output \quad \text{(Equation 9)}$$
$$\{(\alpha \times left\ input) - (\alpha \times right\ input)\} = 0 \equiv left\ output = right\ output \quad \text{(Equation 10)}$$
$$\{(\alpha \times left\ input) - (\alpha \times right\ input)\} < 0 \equiv left\ output < right\ output \quad \text{(Equation 11)}$$

Assuming symmetric sigmoidal activation functions at the output units, it follows from Equations 6-11 that:

$$left\ input > right\ input \equiv left\ output > 0\ and\ right\ output < 0 \quad \text{(Equation 12)}$$
$$left\ input = right\ input \equiv left\ output = 0\ and\ right\ output = 0 \quad \text{(Equation 13)}$$
$$left\ input < right\ input \equiv left\ output < 0\ and\ right\ output > 0 \quad \text{(Equation 14)}$$

Equations 12-14 indicate how the number-comparison decision should be made based on the values of the output units. Importantly, the dual-output model depicted in Figure 10 implies a family of logical models, each of which satisfies the condition that, for each output unit, the sum of the two incoming weights is 0. This logical result generates the additive inverse prediction that is confirmed in our NC neural-network simulations in the previous Results section.

## Discussion

Our NC neural-network-based system simulates a wide range of phenomena: distance and ratio effects and strong generalization to multidigit integers and other, advanced number types (negative numbers and decimal numbers). It is essential to cover distance and ratio effects, showing that the underlying computation is not a memory-based lookup table and supporting a more abstract principle that accuracy and reaction time both improve with numbers that are easier to distinguish, whether by differences or ratios.

This is the first computational model to explore generalization across number types. Previous computational models invariably trained and tested on a single type of number, whether 2-digit integers, negative numbers, or decimal numbers. The NC model uniquely generalizes robustly from training on only the 1-9 digits to multidigit numbers, negative numbers, and decimal numbers, with success rates in the mid .90s. We do not claim children automatically generalize that well to these advanced types because they obviously do not. But we do believe that training on the single digit numbers establishes a strong foundation for moving on to the advanced number types. To generalize well, individuals would also have to learn the technical vocabularies for auditory presentation of number pairs, e.g., millions, billions, trillions, quadrillions, quintillions, sextillions, etc., as well as appropriate scientific notation for written input number presentations.

It is interesting that the strong generalization ability of the NC model provides a notable exception to the claim that artificial neural networks require immense amounts of training and still do not generalize nearly as well as humans do (Lake, Ullman, Tenenbaum, & Gershman, 2017). As an example, we find that NC generalization from training 72 pairs of single digits to a high level of accuracy on over 80 million 4-digit integers is impressively robust. Further study of the reasons for this extraordinary generalization ability could perhaps inspire more efficient and effective human-like AI systems.

An important reason for such strong NC generalization here is that this tiny CC artificial neural network learns the basics of a powerful number system, not merely a large collection of unrelated numerical facts. This powerful system, variously named Arabic, Hindu-Arabic, or Western-Arabic, first emerged in the tenth century BC (1000 BC – 901 BC) (Kunitzsch, 2003; Plofker, 2009). It is a base 10 system, encompassing not only the digits 1-9 but also 0. Importantly, it enables progression to more advanced numeric forms such as multidigit numbers, negative numbers, and decimal numbers. We do not include 0 in our experiments because 0 is very rarely used in psychology experiments on number comparison. However, 0 could readily be included in NC as needed.

Despite harnessing all this potential numeric power, the NC system is considerably simpler than all other existing computational models of number comparison. For example, one of the leading number comparison models employs at least six networks and many parameters and connection weights, and requires up to 100,000 epochs of training (Huber et al., 2016). This higher degree of complexity is designed to deal with fifteen other identified empirical phenomena in number comparison.

Another way to appreciate the relative simplicity of the NC model is to realize that number comparison is essentially a linearly separable problem. A graphical illustration, with $x$ and $y$ representing the two numbers being compared, is presented in Figure 9 as a square plot with a diagonal line where $x = y$. Everywhere above that line $y > x$, while below the line $x > y$. As in other neural networks, NC learning tries to separate and predict the outputs of interest. With only two dimensions, a single straight line does the trick; with three dimensions a plane would be required. More abstractly, networks attempt to carve the input space into however many regions are required by building the correct hyperplane. Training a network to compare the integers 1-9 is sufficient to build the $x = y$ line as the required hyperplane. Generalizing to larger multidigit integers is easily done by moving farther up that fundamental diagonal line. Negative numbers can be accommodated by moving to the lower left below {0, 0}. Decimal numbers are achieved by moving up or down the line in smaller increments.

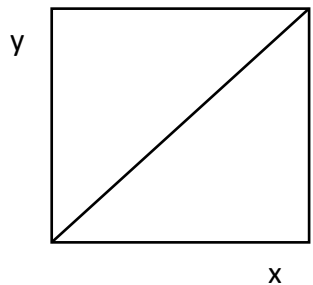

Figure 9. The $x = y$ line readily separates the two outcomes $x > y$ and $y > x$.

Although the number comparison problem is a linearly separable problem, CC can also learn nonlinear problems. CC does this by recruiting hidden units into its network structure, one at a time, as needed (Fahlman & Lebiere, 1990). The greater the nonlinearity, the more hidden units are recruited. In classical CC, recruited hidden units are placed on their own layer, adjacent to the output units. This creates a deep network structure. A popular version of CC, Sibling-descendant Cascade-correlation (SDCC) installs the recruited hidden unit either in this fashion, as a descendant of the last prior recruit, or on the current highest layer, as a sibling of the last prior recruit, depending on which candidate recruit has an activation pattern that correlates best with current network error (Baluja & Fahlman, 1994). This creates network topologies with varying degrees of depth. Yet another member of the CC family, Neural Probability Learner and Sampler (NPLS), can learn probability distributions by stopping the learning process when error reduction stagnates. This is the point at which a probability distribution has been learned (Shultz & Nobandegani, 2021).

Our NC model is so far unique among number comparison models in using integer and sometimes decimal inputs (only for testing on decimal numbers). However, numeric inputs are nearly universal in other artificial neural networks. Numeric inputs are important in starting to learn the Arabic number system, and they are widely used in psychology experiments on number comparison.

An alternative input option used in number comparison experiments is that of dot patterns, presenting a more perceptual task. A popular and effective neural-network technique for coding such perceptual stimuli is *thermometer* coding, in which a number is represented by a set of activated units corresponding to its numerosity (Zorzi & Butterworth, 1999; Zorzi & Testolin, 2018). For example, 6 would be represented by activating units 1-6 from left to right, and 16 by activating units 1-16, again from left to right. Although thermometer coding may appear to be a natural way to represent a collection of dots, it seems implausible that a biological number comparison system would be equipped with precisely the correct number of input units in anticipation of processing a virtually infinite range of number pairs to be compared. We are currently working on a different approach to dot pattern inputs using convolutional deep-learning networks that learn to map perceived object collections onto numbers, avoiding such implausible evolutionary engineering.

Many of the coding techniques used to model number comparison in neural networks share the assumption that larger numbers create higher activation levels than do smaller numbers. This makes sense and works well, particularly by relying on the use of sigmoidal activation functions that effectively convert all numbers to a standard range, most often 0 to 1. This effectively allows a network to benefit from important characteristics of the Arabic number system, including proper subsets, a number line, and compression effects. Smaller numbers are proper subsets of larger numbers. The number line orders the numbers from left to right and seems to compress into smaller steps on the right where numbers get large. As our simulations demonstrate, these two characteristics are either preserved (proper subset) or constructed (number line, compression) in the NC model. Any parts of a number line can be constructed by ordered NC results in a particular number region. Larger small/large ratios can make it seem that far right regions are more compressed than are far left regions. Although a number line and its perceived compressions are often viewed as primary causes of psychological phenomena, NC shows how those constructions could naturally arise during learning and generalization; they do not need to be assumed.

The NC model also provides a relatively simple example of how rules could be implemented in neural systems. Here, a simple feed-forward neural network learns and generalizes with such strong regularity that the outcomes are only distinguishable at the behavioral level with a few small errors. Further study of such simple problems might eventually provide insights into how more complex symbol systems could be implemented in neural systems (Lake et al., 2017). Here, a small NC network generalizes so well from a bit of learning that it only very rarely makes mistakes on more complex, related problems. One example of a symbolic rule system for numbers is our logical model of number comparison in Equations 3-14, designed as a proof for why the NC computational model works. Such symbolic systems could explain error-free performance but would likely have trouble accounting for very rare errors and significant difference and ratio effects, all of which are empirically well established and thus stand as worthy simulation targets for computational models.

We hope that our NC model would inspire empirical research on number comparison in children, including young children. Our modeling was partly inspired by a spontaneous game of number comparison with a grandson at 3 years, 3 months of age. He answered several questions comparing two-digit numbers that he had never encountered before, and he was invariably correct. He also detected our intentional mistakes on 2-digit comparisons that he spontaneously posed to us. More systematic number comparison studies should be done on in children as young as 3 years and older as they move up to larger-digit comparisons and comparing advanced number forms such as such as negative and decimal numbers.

We expect that results of such studies might roughly conform to the first-appearance stages of several documented number-knowledge acquisitions (Siegler, 2022): single-digit integers at 3-5 years, 2-digit integers at 5-7 years, 3-digit integers at 7-12 years, and decimal and negative number forms at 11+ years. Interestingly, there is a representational shift from logarithmic to linear within each of these three age periods. Such a shift is consistent with our view that ratio effects and number-line compressions are emergent mental constructions in humans and neural networks rather than hard-coded features as in the Arabic number system.

It is important to note that our NC model is a neural network which is initially ignorant about the relative magnitudes of its numerical input. The network learns these magnitudes by training on pairs of the digits 1-9. Such training enables very strong, but not perfect, generalization to more advanced tasks involving multi-digit, negative, and decimal numbers. From its performance, a trained network looks as if it is operating at a symbolic level, apart from producing a very few errors. Such error rates are in the neighborhood of those made by adult participants on single-digit pairs (Moyer & Landauer, 1967).

It has also been found that children's discovery of new arithmetic strategies proceeds with few or no flawed strategies (Siegler & Jenkins, 1989). When engaged with a rigorous number system, children's advances are strongly constrained by the system being learned, which is also true in the NC model.

Finally, parents and teachers can do their children a great service by helping them learn to make correct number comparisons of the integers 1-9. This is a case in which a little bit of learning on single-digit comparisons can eventually facilitate powerful reasoning, roughly on par with learning an efficient alphabet in the service of reading acquisition. Children might also be encouraged to play number comparison games with each other as such games are easy to perform in natural settings, with or without visual number symbols, and immensely practical in daily life.